\documentclass{JHEP3}
\usepackage{amssymb}
\usepackage{amsfonts}
\usepackage{amsbsy}
\usepackage{amsmath}
\usepackage{amsthm}
\usepackage{graphicx}
\usepackage{epstopdf}
\usepackage[vcentermath]{youngtab}
\usepackage{multirow}

\newtheorem{claim}{Claim}[section]

\def\Z{\mathbb{Z}}

\def\F{\mathcal{F}}
\def\O{\mathcal{O}}
\def\M{\mathcal{M}}
\def\G{\mathcal{G}}
\def\L{\mathcal{L}}
\def\N{\mathcal{N}}
\def\talpha{\tilde{\alpha}}
\def\nhv{n_h-n_v}
\def\n3a{t}
\def\Tr{{\mathrm{Tr}}}
\def\tr{{\mathrm{tr}}}
\def\half{\frac{1}{2}}
\def\tomega{\tilde{\omega}}

\title{A bound on 6D $\N=1$ supergravities}

\author{Vijay Kumar and Washington Taylor\\
Center for Theoretical Physics\\
Department of Physics\\
Massachusetts Institute of Technology\\
77 Massachusetts Avenue\\
Cambridge, MA 02139, USA\\

{\tt vijayk} {\rm at} {\tt mit.edu},
{\tt wati} {\rm at} {\tt mit.edu}
}

\preprint{MIT-CTP-4077}

\abstract{We prove that there are only finitely many distinct
semi-simple gauge groups and matter representations possible in
consistent 6D chiral $(1,0)$ supergravity theories with one tensor
multiplet.  The proof relies only on features of the low-energy
theory; the consistency conditions we impose are that anomalies
should be cancelled by the Green-Schwarz mechanism, and that the
kinetic terms for all fields should be positive in some region of
moduli space.  This result does not apply to the case of the
non-chiral $(1,1)$ supergravities, which are not constrained by
anomaly cancellation.  }

\begin{document}

\section{Introduction}

The requirement of anomaly cancellation can provide non-trivial
consistency conditions for quantum field theories.  In the context of
the Standard Model, for example, vanishing of gauge anomalies requires
that the number of generations of leptons and quarks are equal.  The
anomalies in any given model are completely determined by the
low-energy effective theory, and in particular, by massless matter
fields that violate parity (chiral fermions, self-dual tensors, etc).
Anomalies arise at one loop and receive no further quantum
corrections; this makes them easy to compute given the matter content
of the model (for a review, see \cite{Harvey-anomalies}).  When the
dimension of space-time is of the form $4k+2$, there are purely
gravitional, purely gauge and mixed gravitional-gauge anomalies
\cite{A-GW}.  In the case of ten dimensional $\N=1$ (chiral)
supergravity coupled to a vector multiplet, it was shown by Green and
Schwarz \cite{Green-Schwarz} that anomalies can be cancelled only when
the gauge group is $SO(32)$, $E_8\times E_8$, $U(1)^{248}\times E_8$
or $U(1)^{496}$.  This is a powerful constraint on this class of
theories.  Our goal here is to explore the analogous constraint in
six-dimensional supergravities.

In this paper, we consider the constraints from anomaly cancellation
in the case of six-dimensional $(1,0)$ supergravity coupled to one
tensor multiplet, and any number of
vector multiplets and matter hypermultiplets.  We
prove the following result --

\begin{center}
\it The set of semi-simple gauge groups and matter representations
appearing in consistent chiral, six-dimensional, $(1,0)$ supergravity
models which have positive gauge-kinetic terms and one tensor multiplet is finite.
\end{center}

We restrict to the case of one tensor multiplet because in this case
 the details of the anomaly
cancellation are simpler, and the theory has a Lagrangian description
.  We have also assumed that the gauge group
contains no $U(1)$ factors.  We expect that a generalization of the
proof given in this paper to include abelian gauge group factors and
an arbitrary number of tensor multiplets may be possible.  Since we
restrict ourselves to {\it chiral} $(1,0)$ supergravity theories, our
conclusions do not apply to models with $(1,1)$ supergravity, which
are always anomaly free.

In Section \ref{sec:review}, we review some properties of the models
we consider and discuss the constraints from anomaly cancellation.  In
Section \ref{sec:finite}, we prove that the set of models is finite.
In Section \ref{sec:classification}, we discuss the possibility of a
systematic classification of allowed models.  In Section
\ref{sec:conclusions}, we summarize and conclude with some comments on
future directions.

\section{Six-dimensional (1,0) supergravity and anomaly cancellation}
\label{sec:review}

We are interested in six-dimensional theories with $(1,0)$
supersymmetry which are gauge theories with matter coupled to gravity
and a single tensor multiplet. The field content of such a theory
consists of single (1, 0) gravity and tensor multiplets, $n_v$ vector
multiplets, and $n_h$ hyper multiplets. Table \ref{table:susy-6d} summarizes the matter content of the $(1,0)$ supersymmetry multiplets we consider in this paper. 
We assume that the gauge
group  $\G=\prod_i G_i$ is semi-simple and contains no $U(1)$ factors.  The low-energy
Lagrangian for such a model takes the form (see \cite{nishino-sezgin,
nishino-sezgin-new, dmw, Seiberg-Witten})
\begin{align}
\L = - \half e^{-2\phi}(dB-\omega)\cdot (dB-\omega)-B\wedge d\tomega
-\sum_i (\alpha_i e^\phi+\talpha_ie^{-\phi}) \tr(F_i^2)+\ldots
\label{eq:lagrangian} 
\end{align}
The index $i$ runs over the various factors in the gauge group.  $\phi$
is the scalar in the tensor multiplet.  Note that there are many terms
in the Lagrangian, including the hypermultiplet kinetic terms,
gravitational couplings and fermion terms, which we have not included
as they are not relevant for our discussion (for details see
\cite{nishino-sezgin, nishino-sezgin-new}).  $\omega,\tomega$ are
Chern-Simons forms defined as
\begin{align}
d\omega = \frac{1}{16\pi^2}\left(\tr R\wedge R-\sum_i \alpha_i \ \tr F_i\wedge F_i\right) \\
d\tomega = \frac{1}{16\pi^2}\left(\tr R\wedge R-\sum_i \talpha_i \ \tr F_i\wedge F_i \right)
\end{align}
The $B$-field transforms under a gauge transformation $\Lambda_i$ in the factor $G_i$ as 
\begin{equation}
\delta B = -\frac{1}{16\pi^2}\sum_i \alpha_i\tr (\Lambda_i F_i)
\end{equation}
The Lagrangian (\ref{eq:lagrangian}) is not gauge invariant, because of
the $B\wedge d\tomega$ coupling; this term is needed for the
six-dimensional analogue \cite{gswest} of the ten-dimensional Green-Schwarz
mechanism\footnote{In the case of six-dimensional models with multiple
tensor multiplets, there is a generalized mechanism due to Sagnotti
\cite{Sagnotti, Ferrara-Minasian-Sagnotti}.  Since we only consider the
case of one tensor multiplet here, this mechanism is not relevant to
the analysis of this paper.}
\cite{Green-Schwarz}.  The tree-level gauge variation of the Lagrangian
cancels the one-loop quantum anomaly, and as a result, the
full quantum effective action is anomaly free.  We  discuss only some
aspects of
this mechanism here,  referring the reader to
\cite{GSW1,GSW2,Polchinski2} for further details.
\begin{table}
\centering
\begin{tabular}{|c|c|}
\hline 
Multiplet & Matter Content \\
\hline
SUGRA & $(g_{\mu \nu}, B^-_{\mu \nu}, \psi^-_{\mu})$ \\
Tensor & $(B^+_{\mu \nu}, \phi, \chi^+)$\\
Vector & $(A_\mu, \lambda^-)$ \\
Hyper & $(4\varphi, \psi^+)$ \\
\hline
\end{tabular}
\caption{Representations of $(1,0)$ supersymmetry in 6D. The + and -
indicate the chirality for fermions and self-duality or
anti-self-duality for the two index tensor.} \label{table:susy-6d}
\end{table}

The gauge anomaly in a $d$-dimensional theory is related to the chiral
anomaly in $d+2$ dimensions, where it is expressed as a $d+2$-form
\cite{Harvey-anomalies}.  Given a six-dimensional $(1,0)$ model with one tensor multiplet, the anomaly polynomial
is an 8-form and takes the form \cite{Erler}
\begin{align}
I &= -\frac{\nhv-244}{5760} \tr R^4 - \frac{44+\nhv}{4608} (\tr R^2)^2 -\frac{1}{96} \tr R^2\sum_i \left[ \Tr F_i^2 - \sum_R x^i_{R} \tr_{R} F_i^2\right] \notag\\
& \quad + \frac{1}{24}\left[ \Tr F_i^4 - \sum_R x^i_{R} \tr_{R}
  F_i^4\right] -\frac{1}{4} \sum_{i,j,R,S} x^{ij}_{RS} (\tr_{R}
F_i^2)(\tr_{S} F_j^2) \label{eq:anomaly-poly} \,,
\end{align}
where $F_i$ denotes the field strength for the simple gauge group
factor $G_i$,  $\tr_{R}$ denotes the trace in representation $R$
of the corresponding gauge group factor, and $\Tr$ denotes the trace in
the adjoint representation.  $x^i_{R}$ and $x^{ij}_{RS}$ denote the numbers of
hypermultiplets in representation $R$ of $G_i$, and $(R,S)$ of
$G_i\times G_j$.    We can express the
traces in terms of the trace in the fundamental representation.
\begin{align}
\tr_{R} F^2 &= A_{R} \tr F^2 \notag \\
\tr_R F^4 &= B_R \tr F^4 + C_R (\tr F^2)^2 \label{eq:ABC-def}
\end{align}
Note that $\tr$ (without any subscript) will always denote the trace
in the fundamental representation.  Formulas for the group-theoretic
coefficients $A_R,B_R,C_R$ can be found in \cite{Erler}\footnote{Note that in \cite{Erler}, the coefficients $v,t,u$ correspond to $A_R,B_R,C_R$ respectively.}.  The
polynomial $I$ can be written in terms of $\tr R^4, \tr F_i^4, \tr
R^2, \tr F_i^2$ using (\ref{eq:ABC-def}).  We rescale the polynomial so
that the coefficient of the $(\tr R^2)^2$ term is one.  Anomalies can
be cancelled through the Green-Schwarz mechanism
when this polynomial can be factorized as
\begin{equation}
I = (\tr R^2 - \sum_i \alpha_i \tr F_i^2)(\tr R^2 - \sum_i \talpha_i \tr F_i^2) \label{eq:factorized-anomaly}
\end{equation}

A necessary condition for the anomaly to factorize in this fashion is
the absence of any irreducible $\tr R^4$ and $\tr F_i^4$ terms.  This
gives the conditions
\begin{align}
\tr R^4   :   & \quad \nhv =  244 \label{eq:grav-anomaly}\\
\tr F_i^4 :   & \quad B^{i}_{Adj}  =  \sum_R x^i_{R} B^i_{R} \label{eq:f4-condition}
\end{align}
For groups $G_i$ which do not have an irreducible $\tr F_i^4$ term,
$B^i_R=0$ for all representations $R$ and therefore
(\ref{eq:f4-condition}) is always satisfied.  The sum in
(\ref{eq:f4-condition}) is over all hypermultiplets that transform
under any representation $R$ of $G_i$.  For example, a single
hypermultiplet that transforms in the representation $(R,S,T)$ of
$G_i\times G_j \times G_k$ contributes $\dim(S)\times \dim(T)$ to
$x^i_R$.

The condition (\ref{eq:grav-anomaly}) plays a key role in controlling
the range of possible anomaly-free theories in 6D.  The anomaly
cancellation conditions constrain the matter transforming under each
gauge group so that the quantity $n_h-n_v$ in general receives a
positive contribution from each gauge group and associated matter, and
the construction of models compatible with (\ref{eq:grav-anomaly})
thus has the flavor of a partition problem.  While there are some
exceptions to this general rule, and some subtleties in this story
when fields are charged under more than one gauge group, this idea
underlies several aspects of the proof of finiteness given here; we
discuss this further in Section \ref{sec:classification}.

After rescaling the anomaly polynomial so that the coefficient of
$(\tr R^2)^2$ is one, when 
(\ref{eq:grav-anomaly}) is satisfied
(\ref{eq:anomaly-poly}) becomes
\begin{align}
I &= (\tr R^2)^2 +\frac{1}{6} \tr R^2\sum_i \left[ \Tr F_i^2 - \sum_R x^i_{R} \tr_{R} F_i^2\right] - \frac{2}{3}\left[ \Tr F_i^4 - \sum_R x^i_{R} \tr_{R} F_i^4\right] \notag\\
& \quad +4 \sum_{i,j,R,S} x^{ij}_{RS} (\tr_{R} F_i^2)(\tr_{S}
F_j^2)
\label{eq:scaled-anomaly}
\end{align}
For a factorization to exist, in addition to (\ref{eq:grav-anomaly}) and 
(\ref{eq:f4-condition}),
the following equations must have a solution for real $\alpha_i, \talpha_i$
\begin{align}
\alpha_i+\talpha_i & = \frac{1}{6} \left( \sum_R x^i_{R} A_{R}^i -
A_{Adj}^i\right)  \label{eq:a-condition}\\
\alpha_i\talpha_i & = \frac{2}{3}\left( \sum_R x^i_{R} C_{R}^i -
C_{Adj}^i \right)  \label{eq:c-condition}\\
\alpha_i\talpha_j +\alpha_j\talpha_i & = 4 \sum_{R,S} x^{ij}_{RS} A_{R}^i A_{S}^j
\end{align}

Notice that the coefficients in the factorized anomaly polynomial
$\alpha_i,\talpha_i$, are related to the coefficients in the
$*B \cdot (F_i\wedge F_i)$ and $B\wedge F_i \wedge F_i$ terms in the Lagrangian
(\ref{eq:lagrangian}).  This was first observed in \cite{Sagnotti}, and
its consequences were discussed further in \cite{dmw,Seiberg-Witten,
Ferrara-Minasian-Sagnotti}.  The coefficients $\alpha_i,
\talpha_i$ are fixed by the anomaly polynomial, which in turn is
determined completely by the choice of the gauge group and the matter
content.  Moreover, due to their origin in the anomaly, the
coefficients $\alpha_i,\talpha_i$ are immune to quantum corrections.
The coefficients $\alpha_i,\talpha_i$ play a key role in the structure
of consistent 6D supergravity theories.  In particular, the signs of
these terms affect the behavior of the gauge fields in the theory at
different values of the dilaton through the gauge kinetic terms in
(\ref{eq:lagrangian}) \cite{Sagnotti}.  When the gauge kinetic term controlled by
$\alpha_i e^\phi+\talpha_ie^{-\phi}$ becomes negative, the theory
develops a perturbative instability.  
When the gauge kinetic term vanishes, the gauge field becomes 
strongly coupled and the theory flows to an interacting 
superconformal field theory in the IR, with
tensionless string excitations \cite{Seiberg-Witten}.  While there are
potentially interesting features associated with some of these
superconformal theories \cite{6D-conformal-fixed-point, intriligator},
in this work we focus on theories where all gauge kinetic terms are
positive for some value of the dilaton.  This imposes the constraint
that there exists some $\phi$ so that
for each gauge group at least one of  $\alpha_i,\talpha_i$ is positive.

As stated in the introduction, the goal of this paper is to
demonstrate that only a {\it finite} number of gauge groups and matter
representations are possible in consistent 6D chiral supergravity
theories with one tensor multiplet.  We prove this in the following
section, based only on the following two assumptions regarding the
theory:

\begin{align}
1.  & \quad \mbox{The anomalies can be cancelled by the Green-Schwarz mechanism.}\nonumber\\
2.  & \quad \mbox{The kinetic terms for all fields are positive in some region of moduli space.} \label{consistency-conditions}
\end{align}

Note in particular that the statement and proof of the finiteness of
this class of theories is purely a statement about the low-energy
effective theory, independent of any explicit realization of a UV
completion of the theories, such as string theory.  Clearly, it is of
interest to understand which low-energy theories admit a consistent UV
completion as quantum gravity theories.  In
\cite{string-universality}, we conjectured that all UV-consistent 6D
supergravity theories can be realized in some limit of string theory.  The
constraints on 6D supergravity theories arising from anomaly
cancellation have intriguing structural similarity to the constraints
associated with specific string compactification mechanisms
\cite{aldazabal, Bianchi, Uranga}.  This correspondence was made explicit for a
particular class of heterotic compactifications in
\cite{string-universality}, and has been explored in the context of
F-theory by Grassi and Morrison \cite{Grassi-Morrison,
Grassi-Morrison-2}, who showed that the anomaly cancellation
conditions in 6D give rise to new nontrivial geometric constraints on
Calabi-Yau compactifications.  The proof of finiteness in this paper
suggests a systematic approach to classifying the complete set of
consistent chiral 6D supergravity theories.  As we discuss in Section
\ref{sec:classification}, such a classification would help in making a
more precise dictionary between the structure of low-energy theories
and string compactifications.

\section{Finiteness of anomaly-free models in 6D}
 \label{sec:finite}

We now proceed to prove the finiteness result stated in the
introduction, using constraints 1 and 2 above.  We will prove
finiteness by contradiction.  Assume there exists an infinite family $
\F $ of distinct models $\M_1,\M_2, \cdots$.  Each model $\M_\gamma$ has
a nonabelian gauge group $\G_\gamma$ which is a product of simple
group factors, and matter hypermultiplets that transform in arbitrary
representations of the gauge group.  The number of hypermultiplets,
the matter representations and the gauge groups themselves are
constrained by anomaly cancellation.  We show that these constraints
are sufficient to demonstrate that no infinite family of anomaly-free
models exists, with distinct combinations of gauge groups and matter.

We first examine some simple consequences of the anomaly
constraints.  The absence of purely gravitational anomalies  requires
that $n_h-n_v+29n_t=273$, and for the case of one tensor multiplet
(\ref{eq:grav-anomaly})
$\nhv=244$.  If  every model in an infinite family $\F$ has the
same gauge group $\G_\gamma=\G$, but with distinct matter
representations, we immediately arrive at a contradiction.  If
$\{R_i\}$ denotes the irreducible representations of $\G$, then we must
have $\sum_i x_i \dim (R_i) = \dim(\G)+244$.  Since $x_i\geq 0 \
\forall i$, and the number of representations of $\G$ below a certain
dimension is bounded, there are only a finite number of
solutions to (\ref{eq:grav-anomaly}).  Therefore, there exists no such infinite family.  If we
assume that model $\M_\gamma$ has a gauge group $\G_\gamma$, such that
the dimension of the gauge group is always bounded from above ($\dim
(\G_\gamma) < M, \ \forall\gamma$), we again arrive at a
contradiction.  This is because there are only finitely many gauge
groups below a certain dimension, and again the gravitational anomaly
condition only has a finite number of solutions.

For any infinite family of anomaly-free models, the dimension of the
gauge group is therefore necessarily 
unbounded.  There are two possibilities for how this can occur, and
we show that in each case the assumption of the existence of an infinite family leads to a contradiction.
\begin{enumerate}
\item The dimension of each simple factor in $\G_\gamma$ is bounded.  In this case, the number of simple factors is unbounded over the family.  

\item The dimension of at least one simple factor in $\G_\gamma$ is unbounded.  For example, the gauge group is of the form $\G_\gamma=SU(N)\times \tilde{G}_\gamma$, where $N\rightarrow \infty$.
\end{enumerate}

\subsection*{Case 1: Bounded simple group factors}

Assume that the dimension of each
simple group factor is bounded from above by $M$.
In this case,
the number of factors in the gauge group necessarily diverges for any infinite
family of distinct models $ \{\M_\gamma\} $.  Any model in the
family has a gauge group of the form $ \G_\gamma=\prod_{k=1}^N G_k $,
where each $G_k$ is simple with $\dim(G_k) \leq M$ and $N$ is
unbounded over the family.

Since each model in the family is free of anomalies, the anomaly
polynomial for each model factorizes as
\begin{equation}
I = (\tr R^2 - \sum_{k=1}^N \alpha_k \tr F_k^2)(\tr R^2 - \sum_{k=1}^N
\talpha_k \tr F^2)
\end{equation}
The positivity condition on the kinetic terms
(\ref{consistency-conditions}) requires that at least one of
$\alpha_k,\talpha_k$ is positive, for each factor $G_k$ in the gauge
group.  The coefficient of $\tr F_i^2\tr F_j^2$ in the anomaly
polynomial (\ref{eq:anomaly-poly}) is related to the number of
hypermultiplets charged simultaneously under both $G_i$ and
$G_j$.  Hence, for every pair of factors $G_i$ and $G_j$ in the gauge
group, the corresponding coefficients in the anomaly polynomial must
satisfy
\begin{equation}
\alpha_i\talpha_j+\alpha_j\talpha_i \geq 0
\end{equation}
From this condition, we infer that {\it at most} one $\alpha_i$ is
negative among all $i$, and at most one $\talpha_j$ is negative among all $j$.  This condition does allow an arbitrary number of $\alpha_i$ or $\talpha_j$ to be zero.  There are three possibilities for a given factor $G_i$ --
\begin{enumerate}
\item {\bf Type 0}: One of $\alpha_i,\talpha_i$ is zero.
\item {\bf Type N}: One of $\alpha_i,\talpha_i$ is negative.
\item {\bf Type P}: Both $\alpha_i,\talpha_i$ are positive.
\end{enumerate}
For any model in the family, there can be at most two type N factors.

We first show that the number of type 0 factors in the gauge group
 is unbounded for any
infinite family.  Assume the contrary is true, that is the number of
such factors is bounded above by $K$ over the entire family.  There
are at most two type N factors, so at least $N-K-2$ factors are of
type P.  For every pair of groups $G_i$ and $G_j$ among these $N-K-2$
factors, $\alpha_i\talpha_j+\alpha_j\talpha_i$ is strictly positive,
so there are matter hypermultiplets charged simultaneously under
$G_i\times G_j$.  
If each of these matter hypermultiplets are charged under at most 
two groups then the number of such hypermultiplets is
$(N-K-2)(N-K-3)/2$, which goes as $\O(N^2)$ for large $N$.  This is an
overcount of the number of hypermultiplets needed, however,
because there could be a
single hypermultiplet charged under $\lambda > 2$ factors, which would
be overcounted $\frac{\lambda(\lambda-1)}{2}$ times.  Let $\lambda$
denote the maximum number of gauge group factors that any
hypermultiplet transforms under.  The dimension of such a
representation $\geq 2^\lambda$.  The number of vector multiplets
scales linearly with $N$, since each factor in the gauge group is
bounded in dimension.  As a result, $\lambda \sim \O(\log N)$; if it
scales faster with $N$, the gravitional anomaly condition $\nhv=244$
would be violated.  In the worst case, if all the matter
hypermultiplets transform under $\lambda \sim \log N$ gauge group factors, $n_h$
still grows as $\O(N^2/\log N)$.    Therefore, $\nhv \sim \O(N^2/\log N)$ as
$N\rightarrow \infty$.  This would clearly violate the gravitational
anomaly condition (\ref{eq:grav-anomaly}) at sufficiently large $N$.

The above argument shows that the number of type 0 factors with
$\alpha$ or $\talpha$ equal to zero must be unbounded.  It also shows
that the number of type P factors with both $\alpha, \talpha$ strictly
positive can grow at most as fast as $\O(\sqrt{N})$ (dropping factors
of $\log N$).  Therefore, there are $\O(N)$ factors of type 0.  For
each factor $G_i$ of this type, the coefficient of the $(\tr F_i^2)^2$
term is zero.  This implies that the third term in the anomaly polynomial 
(\ref{eq:scaled-anomaly}) vanishes for each type 0 factor $G_i$, so
\begin{equation}
\sum_R x_R \tr_R F_i^4 = \Tr F_i^4 \label{eq:special-property}
\end{equation}
where $x_R$ is the number of hypermultiplets in representation $R$ of
the group $G_i$.

We now show that type 0 factors all give a positive contribution to
$n_h-n_v$ in the gravitational anomaly condition
(\ref{eq:grav-anomaly}), so that the number of these factors is bounded.

\begin{claim}
Every gauge group factor $G_i$ that satisfies
(\ref{eq:special-property}) also satisfies $h_i-v_i  > 0$, where
$h_i=\sum_R x_R \dim(R)$ denotes the number of hypermultiplets charged
under $G_i$, and $v_i=\dim(G_i)$.  \label{group-theory-claim}
\end{claim}

The proof of this claim is in Appendix \ref{sec:group-theory}.  Using
the statement of the claim, we wish to show that $\nhv$ is positive
and grows without bound over the family.  We must be careful since
there could be hypermultiplets charged under multiple groups, which
will be overcounted if we simply add $h_i-v_i$ for each factor $G_i$.
It is easily checked that there is no matter charged simultaneously
under three or more type 0 factors\footnote{If there was such a
hypermultiplet charged under $G_1\times G_2\times \cdots \times G_n$,
for $n > 2$, then for all pairs $x_{ij}\propto
\alpha_i\talpha_j+\alpha_j\talpha_i\neq 0$.  This is impossible, since
for at least 2 factors, $\talpha_i=\talpha_j=0$ or
$\alpha_i=\alpha_j=0 \Rightarrow x_{ij}=0$ for every such pair.}
For two factors $G_i, G_j$, hypermultiplets can be
charged under $G_i\times G_j$ if $(\alpha_i,\talpha_i)=(+,0)$ and
$(\alpha_j,\talpha_j)=(0,+)$ (or the other way around).  For large
enough $N$, however,
all the type 0 factors must have either
$\alpha_i=0$ or $\talpha_i=0$, and hence, there cannot
be {\it any} hypermultiplet charged under two type 0 factors $G_i$ and
$G_j$.  To prove this assertion,
assume there is one factor $G_1$ of the form $(0,+)$ and
$\O(N)$ factors are of the other type $(+,0)$.  This implies that the
number of hypermultiplets charged under some representation $R$ of
$G_1$ is large and scales as $\O(N)$.  This is impossible, since this
factor must satisfy $\sum_R x_R \tr_R F_1^4 = \Tr F_1^4$, while the RHS
is fixed and $\tr_R F_1^4 > 0$ for every $R$.  Thus, for large
enough $N$,
\begin{equation}
n_h-n_v = \sum_{\tiny\mbox{type 0}} \left( h_i-v_i\right)+\mbox{P/N-type contribution}
\end{equation}
Since the dimension of each gauge group factor is bounded from above by $M$, the contribution of each factor is bounded from below $\nhv \geq -M$.  So, even if the $\O(\sqrt{N})$ type P factors contributed negatively, 
\begin{equation}
\nhv \geq \O(N) - M\O(\sqrt{N})
\end{equation}
$\nhv$ is positive and grows as $\O(N)$, in contradiction with the gravitional anomaly condition (\ref{eq:grav-anomaly}).

This shows that given a finite list of simple groups, there is no infinite family of anomaly-free models, each of which has a gauge group consisting of an arbitrary product of simple groups in the list and matter in an arbitrary representation.

\subsection*{Case 2: Unbounded simple group factor}

The only other way in which a gauge group could grow unbounded over a
family of models is if the gauge group contains a classical group
$H(N)$ (either $SU(N)$, $SO(N)$ or $Sp(N/2)$), whose rank grows
without bound.  Any such infinite family $\F$ necessarily contains an
infinite sub-family of models with gauge group $H(N)\times G_N$ with $N
\rightarrow \infty$.  
We now show that this case also leads to a contradiction.

\vspace{0.15in} 

\noindent
{\it Brief outline of the proof for case 2}: We examine the $\tr
F^4$ conditions for models in the family with $N > \tilde{N}$.  This
allows us to show that every infinite family with an unbounded factor, at large enough
$N$, has an infinite sub-family with gauge group and matter content parameterized in one of a few different ways.  For
each of these possible parameterizations of gauge group and associated matter fields, the contribution
to $\nhv$ diverges with $N$.  This would violate the gravitational
anomaly condition, unless there is a sufficiently negative
contribution from the rest of the gauge group $G_N$ and the associated
charged matter.  This, however, requires $\dim(G_N)$ to grow without
bound in such a way that the contribution to $\nhv$ is negative and
divergent, which we show is impossible.
\vspace{0.15in}
\begin{table}
\centering
\begin{tabular}{|c|c|c|c|c|c|c|}
\hline
Group & Representation & Dimension & $A_R$ & $B_R$ & $C_R$ \\
\hline
\multirow{7}{*}{$SU(N)$} & ${\tiny\yng(1)}$ & $N$ & 1 & 1 & 0 \\
 & Adjoint & $N^2-1$ & $2N$ & $2N$ & 6 \\
 & ${\tiny\yng(1,1)}$ & $ \frac{N(N-1)}{2} $ & $ N-2 $ & $ N-8 $ & 3 \\
 & ${\tiny\yng(2)}$ & $ \frac{N(N+1)}{2} $ & $ N+2 $ & $ N+8 $ & 3 \\
 & ${\tiny\yng(1,1,1)}$ & $ \frac{N(N-1)(N-2)}{6} $& $ \frac{N^2-5N+6}{2} $&$ \frac{N^2-17N+54}{2} $ & $ 3N-12 $ \\
 & ${\tiny\yng(2,1)}$ & $ \frac{N(N^2-1)}{3} $& $ N^2-3 $& $ N^2-27 $& $ 6N $ \\
 & ${\tiny\yng(3)}$ & $ \frac{N(N+1)(N+2)}{6} $ &$ \frac{N^2+5N+6}{2} $&$ \frac{N^2+17N+54}{2} $ & $ 3N+12 $ \\
\hline
\multirow{3}{*}{$SO(N), Sp(\frac{N}{2})$} & ${\tiny\yng(1)}$ & $N$ & 1 & 1 & 0 \\
 & ${\tiny\yng(1,1)}$ & $\frac{N(N-1)}{2}$ & $N-2$ & $N-8$ & 3 \\
 & ${\tiny\yng(2)}$ & $\frac{N(N+1)}{2}$ & $N+2$ & $N+8$ & 3\\
 \hline
\end{tabular}
\caption{Values of the group-theoretic coefficients $A_R, B_R, C_R$
for some representations of $SU(N)$, $SO(N)$ and $Sp(N/2)$.  Note that
the adjoint representations of $SO(N)$ and $Sp(N/2)$
are given by 
the 2-index antisymmetric and the 2-index symmetric representation
respectively.}
\label{table:unitary-formulas}
\end{table}

Consider first the case when the gauge group is of the form
$SU(N)\times G_N$, where $G_N$ is an arbitrary group.  Anomaly
cancellation (\ref{eq:f4-condition}) for the $SU(N)$ gauge group component
requires that
\begin{eqnarray}
B_{\rm Adj} =2N = \sum_R x_R B_R \label{eq:SU-trace}
\end{eqnarray}
where $R$ is a representation of $SU(N)$.  $x_R$ denotes the total
number of hypermultiplets that transform under representation $R$ of
$SU(N)$; it also includes hypermultiplets that are charged under both
$SU(N)$ and $G_N$.  For example, in a model with gauge group
$SU(N)\times SU(4)$ with matter content $1(N,1)+1(1,6)+2(N,4)+c.c$,
the number of hypermultiplets in the $N$ of $SU(N)$ would be counted
as $x_N=1+2\times 4=9$.  The coefficients $A_R,B_R,C_R$ for various
representations of $SU(N)$ are shown in Table
\ref{table:unitary-formulas}.

The values of $B_R$ for all representations of $SU(N)$ other than the
fundamental, adjoint, and two-index symmetric and antisymmetric
representations grow at least as fast as $\O(N^2)$ as $N\rightarrow
\infty$.  For any given infinite family, there thus exists an
$\tilde{N}$ such that for all $N > \tilde{N}$, the models in the
family have no matter in any representations other than these.  This
follows because the LHS of equation (\ref{eq:SU-trace}) scales as
$\O(N)$, and each $x_R\geq 0$ on the RHS.  Thus, at large enough $N$,
we only need to consider the fundamental, adjoint, symmetric and
anti-symmetric representations.  For these representations,
(\ref{eq:SU-trace}) reads
\begin{eqnarray}
2N = x_1+2N x_2+(N-8)x_3+(N+8)x_4 \label{eq:SU-trace-1}
\end{eqnarray} 

The only solutions to this equation when $N$ is large are shown in
Table \ref{table:solutions}.  We have discarded solutions $(x_1, x_2,
x_3, x_4) = (0, 1, 0, 0)$ and $(0, 0, 1, 1)$, where $\alpha =
\tilde{\alpha}=0$ so the kinetic term for
the gauge field is identically zero.  We repeat the same analysis for
the groups $SO(N)$ and $Sp(N/2)$.  At large values of $N$, we only
need to consider the fundamental, anti-symmetric and symmetric
representations.  The allowed matter hypermultiplets for models with
non-vanishing kinetic term are shown in Table \ref{table:solutions}.
\begin{table}
\centering
	\begin{tabular}{|c|c|c|c|}
	\hline
	Group & Matter content & $\nhv$ & $\alpha,\talpha$ \\
	\hline
	\multirow{4}{*}{$SU(N)$} & $2N\ {\small\yng(1)}$ & $N^2+1$& $2,-2$\\
	& $(N+8)\ {\small\yng(1)}+1\ {\tiny\yng(1,1)}$ & $ \frac{1}{2}N^2+\frac{15}{2}N+1 $& $2,-1$\\
	& $(N-8)\ {\small\yng(1)}+1\ {\small\yng(2)}$ & $ \frac{1}{2}N^2-\frac{15}{2}N+1 $& $-2,1$\\
	& $16\ {\small\yng(1)}+2\ {\tiny\yng(1,1)}$ & $ 15N+1 $ & $2,0$\\
	\hline
	$SO(N)$ & $(N-8)\ {\small\yng(1)}$ & $\frac{1}{2}N^2-\frac{7}{2}N$ & $-2,1$ \\
\hline
\multirow{2}{*}{$Sp(N/2)$} & $(N+8)\ {\small\yng(1)}$ & $\frac{1}{2}N^2+\frac{7}{2}N$ & $2,-1$ \\
	 & $16\ {\small\yng(1)}+1\ {\tiny\yng(1,1)}$ & $15N-1$ & $2,0$ \\
	 \hline
	\end{tabular}
	\caption{Allowed charged matter for an infinite family of models with gauge group $H(N)$.  The last column gives the values of $\alpha,\talpha$ in the factorized anomaly polynomial.}
	\label{table:solutions}
\end{table}

We have shown that for any infinite family of
models where the rank of one of the gauge group factors increases
without bound,
the matter content is restricted to one of those in Table
\ref{table:solutions}.  Notice that for each of these possibilities
the contribution to $\nhv$ from matter charged under $H(N)$ diverges
as $N\rightarrow \infty$ (either as $\O(N)$ or $\O(N^2)$).  Since the
gauge group is $H(N)\times G_N$, the contribution to $\nhv$ from $G_N$
and associated matter charged under this group must be negative and
unbounded (at least $-\O(N)$ or $-\O(N^2)$) in order to satisfy the
gravitational anomaly condition.  If we assume that the dimension of each simple
factor in $G_N$ is bounded, and that the number of simple factors
diverges, we arrive at a contradiction in exactly the same way as we
did in Case 1 above: $\nhv$ is positive and scales at least linearly
with the number of factors.  $G_N$ must therefore contain another
classical group factor $\hat{H}(M)$, whose dimension also increases without
bound over the infinite family.

Therefore, any given infinite family must have an infinite sub-family, with gauge group of the
form $\hat{H}(M)\times H(N) \times \tilde{G}_{M, N}$, with both $M,
N\rightarrow \infty$.  Note that values of $(\alpha,\talpha)$ for each
unbounded classical factor are restricted to the values $(\pm2, \mp
2)$, $(\pm2,\mp1)$, $(\pm1,\mp2)$, $(2,0)$,$(0,2)$.  If $F_1$ denotes the
field strength of the $\hat{H}(M)$ factor and $F_2$ that of $H(N)$, the
coefficient of the $\tr F_1^2\tr F_2^2$ term in the anomaly polynomial
is $4\sum_{R,S} A_RA_S x_{RS}$, where $x_{RS}$ is the number of
hypermultiplets in the $(R,S)$ representation of $\hat{H}(M)\times H(N)$.
Comparing coefficients, we have
\begin{equation}
\alpha_1\talpha_2+\alpha_2\talpha_1 = 4\sum_{R,S} A_R A_S \;x_{RS} \in
4\Z
\end{equation}
From Table \ref{table:solutions}, the coefficients $\alpha_i,\talpha_i$ do not diverge with $N$ and $M$, and therefore the LHS also does not diverge.  The
group-theoretic factors $A_R$ and $A_S$ on the other hand, have
positive leading terms divergent with $N$ and $M$ for all
representations except the fundamental.  Hence, $R$ and $S$ are
restricted to be the fundamental representations of the groups $\hat{H}(M)$
and $H(N)$ respectively.  The RHS of the above equation is, then, just
four times the number of bi-fundamentals and therefore non-negative.
Since at most one $\alpha_i$, and at most one $\talpha_i$
can be negative, we can without loss of generality fix $\alpha_1  >0$
and $\alpha_1 \geq \talpha_2$.
If we require that 
$\alpha_1\talpha_2+\alpha_2\talpha_1$ be divisible by four and
non-negative, the possible values for
$(\alpha_1,\talpha_1,\alpha_2,\talpha_2)$ are --
\begin{enumerate}
\item $(2,-2,0,2)$,$(2,-1,0,2)$,$(2,0,0,2)$: For each of these values,
there is one bi-fundamental hypermultiplet charged under $\hat{H}(M)\times
H(N)$.  As a consequence, the number of hypermultiplets charged under
the fundamental representation of $H(N)$ scales linearly with $M$,
which diverges.  This is in contradiction with Table
\ref{table:solutions} where the number of hypermultiplets in the
fundamental is fixed at 16.

\item $(2,0,2,0)$: In this case, the number of bi-fundamentals is
zero.  The contribution to $\nhv$ from $\hat{H}(M)\times H(N)$
diverges with $M,N$. If such an infinite family were to exist, the
same argument as before implies that $\tilde{G}_{M,N}$ must contain a
classical group factor that grows without bound and cancels the
divergent contribution to $\nhv$. As this analysis shows, this factor
must have $(\alpha,\talpha)=(2,0)$, and we have learned that such
factors contribute a positive, divergent amount to $\nhv$. Therefore,
there is no way that the divergent contribution from $\hat{H}(M)\times
H(N) $ can be compensated by a negative divergence from factors in
$\tilde{G}_{M,N}$ and so we have a contradiction in this case as well.

\item $(2, -2, -2, 2), (2, -1, -2, 1)$: These possibilities give rise
  to five infinite families of models where the anomaly factorizes and
  can be cancelled by the Green-Schwarz mechanism; these families are
  tabulated in Table~\ref{table:infinite-families}.
In all the models in these families, we have $\alpha_1 = -\alpha_2,
\talpha_1 = -\talpha_2$.  Thus, at any given value of the dilaton, the
  sign of the gauge kinetic terms is opposite for the two gauge group
  factors, and most be negative for one gauge group.  Thus, these
  models all have a perturbative instability.  While it is possible
  that there is some way of understanding these tachyonic theories,
  they do not satisfy our condition of positive gauge kinetic terms,
  so we discard them.  Note that the first two of these five infinite
  families were discovered by Schwarz \cite{Schwarz}, and the third
  was found in \cite{string-universality}.  The argument presented
  here shows that this list of five families is comprehensive.
\end{enumerate}

Note that there are no families with three or more gauge groups of
unbounded rank satisfying anomaly factorization.  
The analysis above gives the values allowed of the
$\alpha, \tilde{\alpha}$'s for each possible pair of groups.  With  three
groups, additional bifundamental matter fields between each pair would
be needed so that the matter content for any component group could not
be among the possibilities listed in Table~\ref{table:solutions}.

\begin{table}
\centering
\begin{tabular}{|c|c|c|}
\hline
Gauge Group & Matter content & Anomaly polynomial\\
\hline
$SU(N)\times SU(N)$ & $2({\tiny\yng(1)}, {\tiny\bar{\yng(1)}})$ & $(X-2Y+2Z)(X+2Y-2Z)$\\
\hline
$SO(2N+8)\times Sp(N)$ & $({\tiny\yng(1)},{\tiny\yng(1)})$ & $(X-Y+Z)(X+2Y-2Z)$\\
\hline
$SU(N) \times SO(N+8)$& $({\tiny\yng(1)},{\tiny\yng(1)})+({\tiny\yng(1,1)},1)$ & $(X+Y-Z)(X-2Y+2Z)$\\
\hline
$SU(N) \times SU(N+8)$&
$({\tiny\yng(1)},{\tiny\yng(1)})+({\tiny\yng(1,1)},1)
+(1,{\tiny\yng(2)})$ & $(X-2Y+ 2Z)(X+Y-Z)$\\
\hline
$Sp(N) \times SU(2N+8)$&
$({\tiny\yng(1)},{\tiny\yng(1)})+(1,{\tiny\yng(2)})$ & $(X-2Y+ 2Z)(X+Y-Z)$\\
\hline
\end{tabular} \label{table:infinite-families}
\caption{Infinite families of anomaly-free 6D models, where the anomaly
polynomial factorizes as shown.  $X,Y,Z$ denote $\tr R^2, \tr F_1^2, \tr F_2^2$
respectively, where $F_1$ is the field strength of the first gauge group factor and $F_2$ that of the second. In each of the above models, the number of neutral hypermultiplets is determined from the $\nhv=244$ condition. Note that the $\Box$ of $SU(N)$ can be exchanged for the $\bar{\Box}$, to generate a different model.}
\end{table}

This completes the proof of Case 2, showing that there are no infinite
families containing a simple group factor of unbounded rank.  This in
turn completes the proof that there exists no infinite family of
models with anomalies that can be cancelled by the Green-Schwarz
mechanism and with kinetic terms positive in some region of moduli
space. $\Box$

\section{Classification of models}
\label{sec:classification}

In proving the finiteness of the number of possible gauge fields and
matter representations which are possible in chiral 6D SUSY gauge
theories, we have developed tools which could lead directly to a
systematic enumeration of all possible consistent low-energy models of
this type.  A previous enumeration of some of these models with one
and two gauge group factors was carried out in \cite{Avramis-Kehagias}.

In particular, it is possible to use the gravitational anomaly
and anomaly factorization conditions to analyze possible gauge groups
in a systematic way by considering each simple factor of the gauge
group separately.  The anomaly conditions (\ref{eq:f4-condition}),
(\ref{eq:a-condition}) and (\ref{eq:c-condition}) constrain the
possible sets of matter fields which transform under any given simple
factor in the gauge group, independent of what other factors appear in
the full gauge group.  The gravitational anomaly
(\ref{eq:grav-anomaly}) places a strong limit on the number of
hypermultiplet matter fields which can be included in the theory.
Since, as we found in several places in the proof in the preceding
section, the contribution of components of the gauge group to
$n_h-n_v$ is generally positive, we can think of the problem of
enumerating all possible gauge and matter configurations for chiral 6D
supergravity theories as like a kind of partition problem.  This
problem is complicated by the matter fields charged under more than
one gauge group, which contribute to $n_h$ only once.  Nonetheless, by analyzing individual simple factors and associated
allowed matter representations, we can determine a set of building
blocks from which all chiral 6D theories may be constructed.  The
contribution to $n_h-n_v$ is reasonably large for most possible blocks
(it seems that only a small number of gauge group factors and
associated matter configurations give negative contributions [which
can only appear once] or positive contributions of much less than 30
or 40 to $\nhv$).  Furthermore, the number of bifundamental fields
grows as the square of the number of type P factors in the gauge
group, so it seems that the combinatorial possibilities for combining
blocks are not too vast.  A very crude estimate suggests that the
total number of models may be under a billion, and that a complete
tabulation of all consistent models is probably possible.  Certainly
we do not expect anything like the $\sim 10^{500}$ distinct models
which can arise from 4D type IIB flux compactifications.  It is also
worth mentioning that as $n_t$ increases, the allowed contribution to
$n_h-n_v$ decreases, so that the total number of possibilities may
decrease rapidly for larger $n_t$ despite the more complicated
anomaly-cancellation mechanism.  We will give a more systematic
discussion of the block-based construction of models in a future
paper.

As an example of the limitations on this type of ``building blocks'',
consider $SU(N)$ with $x_f$ hypermultiplet fields in the fundamental
representation and $x_a$ in the antisymmetric two-index
representation.  The condition (\ref{eq:f4-condition}) relates $x_a$
and $x_f$ through $x_f = 2 N -x_a (N -8)$; this relation allows us to
write the contribution to the gravitational anomaly as $n_h-n_v = 1 +
N (x_f + 7x_a)/2$.  This quantity is necessarily positive, and can be
used to bound the allowed values of $N, x_f$, and $x_a$ either for a group with
one factor $SU(N)$ or with several factors of which one is $SU(N)$.  A
similar analysis can be carried out for larger representations of
$SU(N)$.  For example, if we allow the 3-index antisymmetric
representation ${\tiny \yng(1,1,1)}$, with a gauge group $U(N)$ and no
additional factors, we find that no model has matter in the
representation ${\tiny \yng(1,1,1)}$ unless $N \leq 8$.  There are
some examples of this type of model, such as the SU(7) theory with
matter $24\ {\tiny \yng(1)} + 2\ {\tiny \yng(1,1)} +{\tiny \yng(1,1,1)}$,
which satisfy anomaly cancellation and which are (we believe) not yet
identified as string compactifications \footnote{As this work was
being completed we learned that Grassi and Morrison have found a local
construction of this kind of
3-index antisymmetric representation through
string compactification in the language of F-theory
\cite{Grassi-Morrison-2}}.

A complete classification of allowed 6D theories would have a
number of potential applications.  In analogy to the story in 10D,
where such a classification led to the discovery of the heterotic $E_8
\times E_8$ string, discovery of novel consistent low-energy models in six
dimensions may suggest new mechanisms of string compactification.  Or,
finding a set of apparently consistent theories which do not have
string realizations might lead either to a discovery of new
consistency conditions which need to be imposed on the low-energy
theory, or to a clearer understanding of a 6D ``swampland''
\cite{vafa-swampland, swampland-2} of apparently consistent theories
not realized in string theory.  If all theories satisfying the
constraints we are using here
can be either definitively realized as string compactifications or
shown to be inconsistent, it would prove the conjecture of ``string
universality'' stated in \cite{string-universality} for chiral 6D
supersymmetric theories, at least for the class of models with one
tensor multiplet and no $U(1)$ gauge factors.

The proof given here has shown that there are a finite number of
distinct gauge groups and matter content which can be realized in
chiral 6D supergravity theories with one tensor multiplet.  We have
not, however, addressed the question of whether a given gauge group
and matter content can be associated with more than one UV-complete
supergravity theory (by ``theory'' here, meaning really a continuous
component of moduli space).  In \cite{Kumar-Taylor}, we showed that
for one class of gauge groups, almost all anomaly-free matter
configurations are realized in a unique fashion in heterotic
compactifications, but that some models can be realized in distinct
fashions characterized by topological invariants described in that
case by the structure of a lattice embedding.  It would be interesting
to understand more generally the extent to which the models considered
here have unique UV completions.

A complete classification of allowed theories in six dimensions could
lead to a better understanding of how various classes of string
compactifications populate the string landscape.  Such lessons might
be helpful in understanding the more complicated case of
four-dimensional field theories with gravity.  As mentioned in Section
\ref{sec:review}, in some situations the anomaly constraints in six
dimensions correspond precisely to the constraints on string
compactifications; in both cases these constraints follow from various
index theorems.  Making this connection more precise in six dimensions
could shed new light on the relationship between string theory and
low-energy effective theories in any dimension.

\section{Conclusions and future directions}
\label{sec:conclusions}

We have shown that for 6D $(1,0)$ supersymmetric theories with
gravity, one tensor multiplet, a semi-simple gauge group and
hypermultiplet matter in an arbitrary representation, the conditions
from anomaly cancellation and positivity of the kinetic terms suffice
to prove finiteness of the set of possible gauge groups and matter content.

In this analysis we have only considered semi-simple gauge groups.
When there are $U(1)$ factors in the gauge group, there is a
generalized Green-Schwarz mechanism discussed in \cite{berkooz}, which
involves the tree-level exchange of a 0-form.  Addressing the question
of finiteness of theories including $U(1)$ factors would require further
analysis.

We have also restricted attention here to theories with one tensor
multiplet, which admit a low-energy Lagrangian description
\cite{marcus-schwarz}.  There are many string compactifications which
give rise to 6D models with more than one tensor multiplet
\cite{Sagnotti-others, Sen-K3, Gimon-Johnson, Dabholkar, 
multiple-tensors}. While a proof of finiteness for models with more
tensor multiplets is probably possible, the exchange of multiple
anti-self-dual tensor fields in the Green-Schwarz mechanism as
described by Sagnotti \cite{Sagnotti} makes this analysis more
complicated, and we leave a treatment of such cases to future work.

A comment may also be helpful on non-chiral theories with (1, 1)
supersymmetry; this issue is addressed in further detail in Appendix
\ref{sec:adjoint-hypermultiplet}.  A $(1,1)$ model in six dimensions
also has $(1,0)$ supersymmetry, but contains an additional (1, 0)
gravitino multiplet beyond the gravity, tensor, hyper and vector (1,
0) multiplets in the theories we have considered here.  It seems
that one
cannot
include the gravitino multiplet without having $(1,1)$ local
supersymmetry, and we have restricted our attention here to models
with $(1,0)$ supersymmetry without this gravitino multiplet.  Since $(1,1)$
models are non-chiral, they cannot be constrained by anomaly
cancellation.  Some further mechanism would be needed to constrain the
set of (1, 1) supersymmetric models in six dimensions.  
It is possible that such constraints could be found by demonstrating
that string-like excitations of the theory charged under the tensor
field must be included in the quantum theory; anomalies in the
world-volume theory of these strings would then impose constraints on
the 6D bulk theory, as suggested in \cite{Uranga}.  It may also be
that understanding the dictionary between anomaly constraints and
constraints arising from string compactification for chiral theories
may suggest a new set of constraints even for non-chiral 6D theories.

The result of this paper ties into the question of the number of
topologically distinct Calabi-Yau manifolds, since 6D supergravity
theories can be realized by compactification of F-theory on
elliptically fibered Calabi-Yau manifolds \cite{F-theory}.
It has been shown by Gross \cite{Gross-finite} that there are only finitely
many Calabi-Yau manifolds that admit an elliptic fibration, up to
birational equivalence.  If we can prove that the set of $(1,0)$
models with any number of tensor multiplets is finite then
this would constitute a
``physics proof'' of the theorem.

The total number of consistent 6D models of the type we consider does
not seem to be enormous.  It is not hard to imagine that these
theories could be completely enumerated in a systematic manner.  This
programme would be very useful to understand the structure of the
landscape and swampland \cite{vafa-swampland} in this special case of
six dimensions with $(1,0)$ supersymmetry.  In analogy with the
discovery of the $E_8\times E_8$ heterotic string, we are optimistic
that further study of the set of consistent 6D supergravity theories
will help us better understand the rich structure of string
compactifications, perhaps giving lessons which will be relevant to
the more challenging case of compactifications to four dimensions.

\vspace{0.15in}

\parindent 0in

{\bf Acknowledgements}: We would like to thank Allan Adams, Michael
Douglas, Daniel Freedman, Ken Intriligator, John McGreevy, and David Morrison for helpful discussions.
This research was supported by the DOE
under contract \#DE-FC02-94ER40818.

\parindent 0.28in

\appendix

\section{Proof of Claim 3.1}
\label{sec:group-theory}

In the proof of finiteness in Section \ref{sec:finite}, we claimed that for any group $G$ if 
\begin{equation}
\sum_R x_R \tr_R F^4 = \Tr F^4, \label{eq:trace-condition}
\end{equation}
for $x_R\in \Z, x_R \geq 0$ and $R$ runs over all the representations of $G$, then
\begin{equation}
h - v = \sum_R x_R \dim(R) - \dim(G) \geq c > 0 \label{eq:pos-condition}
\end{equation}

Equation (\ref{eq:trace-condition}) is automatically satisfied if
$x_R=1$ for the adjoint representation and 0 for all other
representations.  In this case, however, the kinetic terms for the
gauge group factor $G$ would be zero, and we are not considering this
case (See Appendix \ref{sec:adjoint-hypermultiplet} for a discussion
of this situation in the context of $(1, 1)$ supersymmetry).  If $x_R
> 0$ for a representation $R$ with $\dim (R) > \dim(G)$, equation
(\ref{eq:pos-condition}) is automatically satisfied.  We only need to
consider situations where $x_R = 0$ for all representations $R$ with
$\dim(R) > \dim(G)$.

\subsection{Exceptional groups}

We first consider the exceptional groups $G_2,F_4,E_6,E_7,E_8$.  The only 
representation that satisfies $\dim(R) \leq \dim(G)$ in all these cases is the fundamental.  Equation (\ref{eq:trace-condition}) is satisfied for these groups, if the number of hypermultiplets are --
\begin{itemize}
\item $G_2$: $x_f=10\Rightarrow h-v = 70-14=56 > 0$
\item $F_4$: $x_f=5\Rightarrow h-v =  130-52=78> 0$
\item $E_6$: $x_f=6\Rightarrow h-v =  162-78=84> 0$
\item $E_7$: $x_f=4\Rightarrow h-v =  224-133=91> 0$
\item $E_8$: There are no solutions because the fundamental representation of $E_8$ is the adjoint representation.
\end{itemize}

\subsection{$SU(N)$}

We first consider the case when $N \geq 4$, and then the cases $SU(2)$ and $SU(3)$. 
The set of irreducible representations are in one-one correspondence with the set of Young diagrams with up to $N-1$ rows.  Equation (\ref{eq:trace-condition}) implies that
\begin{equation}
\sum_R x_R C_R = 6 \label{eq:C-condition}
\end{equation}
Since $x_R \geq 0$, we only need to consider representations which have $C_R \leq 6$, where the coefficients $C_R$ are defined in (\ref{eq:ABC-def}). 
For a given representation $R$ of $SU(N)$, choose $F=F_{12} T^{12}_R + F_{34} T^{34}_R$.  The generators $T^{12}$ and $T^{34}$ are given (in the fundamental representation) by
\begin{eqnarray}
(T^{12})_{ab} & = & \delta_{a1}\delta_{b1} - \delta_{a2}\delta_{b2}  \\
(T^{34})_{ab} & = & \delta_{a3}\delta_{b3} - \delta_{a4}\delta_{b4} , \ a,b=1,2,\cdots ,N 
\end{eqnarray}
Substituting this form of $F$ into the definition (\ref{eq:ABC-def}) for $C_R$, we have
\begin{eqnarray}
\tr_R (F_{12}T^{12}_R+F_{34}T^{34}_R)^4 & = & (2B_R+4C_R)(F_{12}^4+F_{34}^4) + 8C_RF_{12}^2F_{34}^2 
\end{eqnarray}
Comparing coefficients of $F_{12}^2F_{34}^2$ on both sides, we have the following formula for $C_R$ --
\begin{equation}
C_R = \frac{3}{4} \tr_R [(T^{12}_R)^2(T^{34}_R)^2] \label{eq:C-formula}
\end{equation}
Using the above formular for $C_R$, we can show that the only
representations of $SU(N)$ that satisfy $C_R \leq 6$ are
\begin{equation}
\mbox{Adjoint, } \ { \yng(1)}, \ { \yng(2)}, \ { \yng(1,1)} \mbox{ for
  all }N, \mbox{ and } { \yng(1,1,1)} \ 
(C_R = 6\mbox{ for }N=6) \label{eq:su-solutions}
\end{equation}

\parindent 0in
{\bf Examples}:
\begin{enumerate}
\item {\footnotesize\yng(2)}: We must compute the trace in (\ref{eq:C-formula}).  The only states that give a non-zero contribution are 
\begin{equation*}
\young(13), \ \young(23), \ \young(14), \ \young(24)
\end{equation*}
This gives $C = 3$ for the two-index symmetric representation.

\item {$\footnotesize\yng(1,1)$}: The only states that contribute to the trace are --
\begin{equation}
\young(1,3), \ \young(2,3), \ \young(1,4), \ \young(2,4)
\end{equation}
Again, here $C = 3$

\item {$\footnotesize\yng(2,1)$}: The following states contribute to the trace --
\begin{eqnarray*}
& \young(13,i), \ \young(24,i), \ \young(14,i), \ \young(23,i), \\
& \young(1i,3), \ \young(2i,4), \ \young(1i,4), \ \young(2i,3), \\
& \young(11,3), \ \young(11,4), \young(22,3), \ \young(22,4), \ \young(33,1), \ \young(33,2), \ \young(44,1), \ \young(44,2)
\end{eqnarray*}
Here $i$ denotes any of the remaining $N-4$ indices.  For this representation 
\begin{equation}
C = \frac{3}{4}(4(N-4)+4(N-4)+4\times 8)=6N
\end{equation}
\end{enumerate}

\parindent 0.28in

Any representation of $SU(N)$ can be represented as a Young diagram
with at most $N-1$ rows. For a general Young diagram, which does not
correspond to the representations listed in (\ref{eq:su-solutions}),
we can show that $C_R > 6$, by explicitly enumerating states that
contribute to $C_R$ as in the examples above.  
In proving this generally, it is useful to note that adding boxes to
any nonempty horizontal row only increases the value of $C$, so it is
sufficient to consider only the totally antisymmetric representations
and case 3 above.

The only solutions to
(\ref{eq:trace-condition}) are --
\begin{enumerate}
\item $1\ {\tiny\yng(1,1)}+1\ {\tiny \yng(2)}$: $h-v=1>0$.  In this case $\alpha=\talpha=0$, so we discard this solution.  
\item $16\ {\tiny\yng(1)}+2\ {\tiny \yng(1,1)}$: $h-v=15N+1>0$
\item $1\ {\tiny\yng(1,1,1)}+24\ {\tiny\yng(1)}, \ (N=6)$: $h-v=20+144-35=129>0$
\end{enumerate}

Now consider the case of $SU(2)$. Since $B_R=0$, (\ref{eq:trace-condition}) for $SU(2)$ becomes
\begin{equation}
\sum_R x_R C_R = 8 \label{eq:SU2-trace}
\end{equation}
We therefore must enumerate all representations where $C_R \leq 8$. The formula for $C_R$ is simpler in the $SU(2)$ case ---
\begin{equation}
C_R = \frac{1}{4} \tr_R [(T^{12}_R)^4] \label{eq:SU2-C-def}
\end{equation}
Young diagrams for $SU(2)$ only have one row. For a diagram with $c$
columns, the states ${\small \young(11\cdots1)}$ and ${\small
\young(22\cdots2)}$ show that $C_R \geq c^4/2 > 8$ for $c > 2$. 
For $c = 2$, we have the adjoint representation with $C_R = 8$, which
would lead to $\alpha = \tilde{\alpha} = 0$.
The
only possible remaining solution to (\ref{eq:SU2-trace}) is $16 \ {\small
\yng(1)}$ hypermultiplets, which has $h-v=29>0$.

For $SU(3)$, (\ref{eq:trace-condition}) becomes
\begin{equation}
\sum_R x_R C_R = 9 \label{eq:SU3-trace}
\end{equation}
We can compute $C_R$ using formula (\ref{eq:SU2-C-def}) for the $SU(3)$ case as well. The Young diagrams for $SU(3)$ contain at most two rows. For a diagram with $c$ columns in the first row, the following states
\begin{equation}
\young(11\cdots1,3\cdots3) \quad \young(22\cdots2,3\cdots3)
\end{equation}
give a lower bound $C_R \geq c^4/2$, with $C_R > 9$ for $c >2$. The
only representations (besides the adjoint) with $C_R \leq 9$ are
${\small \yng(1), \ \yng(2)}$ with $C_R=1/2, \ 17/2$. The only
combination of  matter representations that satisfies (\ref{eq:SU3-trace}) with one of $\alpha,\talpha$ positive is $18\ {\small \yng(1)}$ with $h-v=46>0$.

\subsection{$SO(N)$}

For $N \leq 6$, these are related to other simple Lie groups, so we only consider $N\geq 7$.  We choose the commuting generators $T^{12}$ and $T^{34}$ of $SO(N)$, defined as
\begin{align}
(T^{12})_{ab} & = i\delta_{a2}\delta_{b1}-i\delta_{a1}\delta_{b2} \\
(T^{34})_{ab} & = i\delta_{a4}\delta_{b3}-i\delta_{a3}\delta_{b4}
\end{align}
Notice that the squared $SO(N)$ generators are identical to the squared $SU(N)$ generators we used in the previous section.  Thus, the formulae for any Young diagram carry over.  In the case of $SO(N)$, the Young diagrams are restricted so that the total number of boxes in the first two columns does not exceed $N$ \cite{Hamermesh}.  The antisymmetric representation is the adjoint, and so we need to find all diagrams with $C_R \leq 3$.  The only diagrams that satisfy this requirement are -- fundamental, 2-index antisymmetric and the 2-index symmetric representation.  

In addition to these, we also have the spinor representations of $SO(N)$.  The irreducible spinor representation of $SO(N)$ has dimension $2^{\lfloor (N-1)/2 \rfloor}$.  It is smaller than the adjoint only for $N \leq 14$, and it can be checked (using the tables in \cite{Patera}) that the only spinor representation that is smaller in dimension than the adjoint is the Weyl/Dirac spinor.  The trace formula for these is \cite{Erler}
\begin{align}
\tr_s F^2 & = 2^{\lfloor (N+1)/2 \rfloor -4} \tr F^2 \\
\tr_s F^4 & = -2^{\lfloor (N+1)/2 \rfloor - 5} \tr F^4 + 3 \cdot 2^{\lfloor (N+1)/2\rfloor -7} (\tr F^2)^2
\end{align}
For the spinor representation $C \leq 3$ for $7\leq N \leq 14$.  For each of these cases, we can solve for representations that solve (\ref{eq:trace-condition}), and check whether $h-v$ is positive.  This is the case for all the solutions, and these are shown in Table \ref{table:SO-solutions}.

\begin{table}
\centering
\begin{tabular}{|c|c|c|}
\hline
$N$ & Matter & $h-v$ \\
\hline
$7$& $8 \mbox{ spinor }+3 \ {\footnotesize\yng(1)}$& $64+21-21=64>0$ \\
$8$& $8 \mbox{ spinor }+4 \ {\footnotesize\yng(1)}$& $64+32-28=68>0$ \\
$9$& $4 \mbox{ spinor }+ 5\ {\footnotesize\yng(1)}$& $64+45-36=73>0$ \\
$10$& $4 \mbox{ spinor }+6\ {\footnotesize\yng(1)}$& $64+60-45=79>0$ \\
$11$& $2 \mbox{ spinor }+7\ {\footnotesize\yng(1)}$& $64+77-55=86>0$ \\
$12$& $2 \mbox{ spinor }+8\ {\footnotesize\yng(1)}$& $64+96-66=94>0$ \\
$13$& $1 \mbox{ spinor }+9\ {\footnotesize\yng(1)}$& $64+117-78=113>0$ \\
$14$& $1 \mbox{ spinor }+10\ {\footnotesize\yng(1)}$& $64+126-91=99>0$ \\
\hline
\end{tabular}
\caption{Matter hypermultiplets that solve (\protect \ref{eq:trace-condition}) for $SO(N)$, $7\leq N\leq 14$.  All solutions for $N > 14$ have positive $h-v$.} \label{table:SO-solutions}
\end{table}

\subsection{$Sp(N)$}

By $Sp(N)$, we mean the group of $2N\times 2N$ matrices that preserve a non-degenerate, skew-symmetric bilinear form.  We choose as generators $T^{12}$, $T^{34}$ in the fundamental representation as
\begin{align}
(T^{12})_{ab} & = \delta_{a1}\delta_{b2}+\delta_{a2}\delta_{b1} \\
(T^{34})_{ab} & = \delta_{a3}\delta_{b4}+\delta_{a4}\delta_{b3}, \quad a,b=1,2\cdots 2N
\end{align}
Again, these generators have been chosen so that their squares are
equal to the squares of the generators of $SU(2N)$.  The Young
diagrams for $Sp(N)$ are similar to those of $SU(2N)$, except that
only diagrams with less than or equal to $N$ rows need to be
considered \cite{Hamermesh}.  The adjoint of $Sp(N)$ is the symmetric
representation.  The analysis for the $SU(N)$ case carries through,
except that we only need to consider representations with $C \leq 3$.
The only representations with this property are -- fundamental,
2-index antisymmetric (traceless w.r.t skew-symmetric form) and the
2-index symmetric.  The only solution is $1 {\tiny\yng(1,1)}+16
{\tiny\yng(1)}$, and for this solution, $h-v=30N+1>0$.  This proves
Claim \ref{group-theory-claim} in Section \ref{sec:finite}.  \hfill
$\Box$

\section{$(1, 1)$ supersymmetry and
the case of one adjoint hypermultiplet}
\label{sec:adjoint-hypermultiplet}

In this paper we have focused on $(1, 0)$ theories with chiral matter
content.  In this appendix we discuss how these theories differ from
$(1, 1)$ non-chiral 6D supergravity theories \footnote{Thanks to Ken
  Intriligator for discussions on this issue.}.

One can imagine a $(1, 0)$ theory where the matter content consists of
precisely one hypermultiplet in the adjoint representation for each
factor in the gauge group.  Together, these $(1,0)$ multiplets combine
to form the $(1,1)$ vector multiplet.  The field content of this
multiplet is \cite{Polchinski2}
\begin{equation*}
A_\mu+4\phi+\psi_+ +\psi_-
\end{equation*}
Since this matter content is non-chiral, there are no gauge
anomalies or mixed gravitational-gauge anomalies; there is still a
purely gravitational anomaly, which is cancelled in the usual way.
In the framework of the discussion here, (\ref{eq:f4-condition}),
(\ref{eq:a-condition}), and (\ref{eq:c-condition}) are all satisfied
with $\alpha_i = \tilde{\alpha}_i = 0$.  Thus, the gauge kinetic terms
vanish and we do not consider models of this type in the analysis
here.

It may seem that this contradicts the straightforward observation that
all $(1, 1)$ supergravity theories can be thought of as $(1, 0)$
gravity theories while the gauge fields in $(1, 1)$
supergravity theories can have nonvanishing gauge kinetic terms.  The
point, however, is that the model described above with adjoint matter
for each gauge group component does not have $(1,1)$ (local) supersymmetry.  To
see this, note that the $(1,1)$ gravity multiplet consists of
\begin{align*}
 & g_{\mu\nu}+B_{\mu\nu}^+ +B_{\mu\nu}^-+4A_\mu+\phi+\psi_\mu^++\psi_\mu^-+\chi_++\chi_- \\
=& (g_{\mu\nu}+B_{\mu\nu}^-+\psi_\mu^-) + (B_{\mu\nu}^++\phi+\chi^+)+(4A_\mu+\psi_\mu^++\chi^-)
\end{align*}
In the $(1,0)$ language, this corresponds to the gravity multiplet,
tensor multiplet and an additional gravitino multiplet.  The $(1, 0)$
gravitino multiplet consists of four abelian vectors, two Weyl
fermions and one gravitino, and is scarcely even mentioned in the vast
literature on 6D models.  The reason for this is that if we consider a
$(1,0)$ model with one gravitino multiplet, it seems that
the model must have
$(1,1)$ supergravity.  This is certainly the case in string theories,
where the vertex operator for the supercharge is the same as that of
the gravitino.  More generally, the common lore \cite{GSW2} states
that a massless, interacting spin-3/2 field must couple to a local
conserved supercurrent.  This fact is compatible with the anomaly
conditions.  The anomaly polynomial in the case of one gravitino
multiplet is
\begin{align}
I &= -\frac{\nhv}{5760} \tr R^4 - \frac{\nhv}{4608} (\tr R^2)^2 -\frac{1}{96} \tr R^2\sum_i \left[ \Tr F_i^2 - \sum_R x^i_{R} \tr_{R} F_i^2\right] \notag\\
& \quad + \frac{1}{24}\left[ \Tr F_i^4 - \sum_R x^i_{R} \tr_{R} F_i^4\right] -\frac{1}{4} \sum_{i,j,R,S} x^{ij}_{RS} (\tr_{R} F_i^2)(\tr_{S} F_j^2)
\end{align}
The gravitational anomaly can be cancelled only if $\nhv=0$.  One way
to cancel the anomaly is to have a single hypermultiplet transforming
in the adjoint of the gauge group $\G$.  In this case, the matter
content is that of a $(1,1)$ 6D model with gravity and one vector
multiplet.  It would be nice to have a simple proof directly from the
anomaly cancellation conditions that this is the only way to have
$\nhv = 0$, which would amount to a proof that any $(1, 0)$ theory
with a gravitino multiplet would have $(1, 1)$ supersymmetry.  In the
case of two gravitino multiplets, it seems that
the only consistent solution is the non-chiral,  
maximal $(2,2)$ 6D supergravity.

We can now understand the apparent discrepancy alluded to above.  A
$(1,0)$ model with one hypermultiplet in the adjoint of the gauge
group $\G$ has $\alpha=\talpha=0$, which implies that the gauge
kinetic terms are zero.  However, in a $(1,1)$ model with a vector
multiplet corresponding to $\G$, the gauge kinetic term is positive.
(An easy way to see this is to consider a $T^4$ compactification of
the heterotic string, which gives $(1,1)$ supergravity in 6D).  This
apparent inconsistency is due to the fact that in the case of the
$(1,1)$ theory, $\alpha\neq 0, \talpha=0$, and this is sufficient to
ensure that the anomaly polynomial vanishes.  The $B$-field has the
usual Chern-Simons coupling that makes it transform under gauge
transformations.  However, there is no $B\wedge d\tomega$ term; the
theory is non-anomalous and therefore there is no need for a
Green-Schwarz counterterm.

Thus, we see that the conditions we impose in this paper on the field
content of the $(1, 0)$ theory exclude the gravitino multiplet
needed to complete the $(1, 0)$ graviton multiplet to a $(1, 1)$
graviton multiplet.  As a result, the class of theories considered
here does not include theories with $(1, 1)$ supersymmetry; the $(1,
1)$ theories are all anomaly free and would need to be constrained by
some alternative mechanism as discussed in \ref{sec:conclusions}.

\end{document}